\documentclass[12pt]{article}
\input epsfig.sty
\usepackage{graphicx} 
\def\laq{\raise 0.4ex\hbox{$<$}\kern -0.8em\lower 0.62
ex\hbox{$\sim$}}
\def\gaq{\raise 0.4ex\hbox{$>$}\kern -0.7em\lower 0.62
ex\hbox{$\sim$}}

\begin{document}

\begin{titlepage}
\begin{flushright}
CERN-PH-TH/2007-114
\end{flushright}
\vspace*{1cm}

\begin{center}
{\LARGE {\bf Large-scale magnetic fields, curvature fluctuations \\
\vskip0.5cm
and the thermal history of the Universe}}
\vskip2.cm
\large{Massimo Giovannini \footnote{e-mail address: massimo.giovannini@cern.ch}}
\vskip 1.cm
{\it Centro ``Enrico Fermi",  Via Panisperna 89/A, 00184 Rome, Italy}
\vskip 0.5cm
{\it Department of Physics, Theory Division, CERN, 1211 Geneva 23, Switzerland}

\end{center}
\begin{abstract}
It is shown that gravitating magnetic fields affect the evolution 
of curvature perturbations in a way that is reminiscent of a pristine non-adiabatic pressure fluctuation.
The gauge-invariant evolution of curvature perturbations is used to constrain the magnetic power spectrum.
Depending on the essential features of the thermodynamic history of the Universe, the explicit derivation of the bound is modified.
The theoretical uncertainty in the constraints on the magnetic energy 
spectrum is assessed by comparing the results obtained in the case of the 
conventional thermal history with the estimates stemming from less
conventional (but phenomenologically allowed) post-inflationary 
evolutions.
\end{abstract}
\end{titlepage}

\newpage
\renewcommand{\theequation}{1.\arabic{equation}}
\setcounter{equation}{0}
\section{Introduction}
\label{sec1}
It is tempting to speculate that large-scale magnetic fields are 
generated in the early Universe \cite{zeldovich1,battaner,kro} during a phase where 
the Weyl invariance of their evolution equations is 
broken either spontaneously of explicitly (see, for instance,  
\cite{maxg1} and references therein). 
As a result the obtained magnetic fields will be necessarily 
tangled over typical wavelengths larger than the Hubble radius
\cite{rat,tw,mgv,mgint,kerstin}. 
In this situation, the presence of magnetic fields may affect 
the evolution of curvature inhomogeneities 
in a way that is reminiscent of what happens in the presence 
of a non-adiabatic pressure fluctuations.

Suppose, indeed, that large-scale magnetic  fields are generated
thanks to the amplification of the quantum fluctuations of an Abelian
gauge field which can be identified with the hypercharge field whose 
modes will not be screened by thermal effects. In spite of the 
specific model of parametric amplification, the result of the process will be a stochastically 
distributed  field whose two point function can be written as 
\begin{equation}
\langle B_{i}(\vec{k},\tau) B^{j}(\vec{p},\tau)\rangle = 
\frac{2\pi^2}{k^3} P_{i}^{j}(k) P_{\mathrm{B}}(k) \delta^{(3)}(\vec{k} + \vec{p}),
\label{corr}
\end{equation}
where $P_{\mathrm{B}}(k)$ is the magnetic power spectrum (which 
may change depending upon the specific way the Weyl invariance is broken)
and $P_{i}^{j}(k) = (\delta_{i}^{j} - k_{i}k^{j}/k^2)$ is the traceless projector.
In Eq. (\ref{corr}) the momenta appearing in the correlator are 
the comoving wave-numbers.
The homogeneous component of this configuration vanishes 
and, as a result, the magnetic fields will not break spatial isotropy
\footnote{If a background magnetic field exist, it must necessarily be oriented along a specific direction \cite{iso1}. The effective  
spatial isotropy is broken and  the angular power spectrum depends 
 on the direction of the magnetic field. As a consequence a number of 
 constraints, stemming directly from the breaking of spatial isotropy, can be derived \cite{iso2}.} of the
 background geometry whose line element can be written, in the spatially flat case and in terms 
 of the conformal time coordinate $\tau$, as
\begin{equation}
ds^2 = a^2(\tau) [ d\tau^2 - d\vec{x}^2].
\label{linel}
\end{equation}
The typical wavelengths that set the initial conditions of the CMB anisotropies \footnote{By this expression
we mean the wavelengths that set the initial conditions of the lowest multipoles 
of the Boltzmann hierarchy.}
are still larger than the Hubble radius (i.e. $k \tau <1$ ) at recombination, taking place 
for an approximate redshift $z_{\mathrm{rec}} \sim 1050$ when the visibility function 
is peaked. 
When $k\tau <1$ the evolution equations of the perturbations of the 
spatial curvature will receive contributions from three independent 
sources. The adiabatic mode typically provides the first contribution
arising from the quantum fluctuations of a (single) inflaton field. 
The second contribution may come from a collection of non-adiabatic 
modes that may be present either because the inflationary evolution is 
driven by more than one scalar degree of freedom, or because 
the primordial plasma contains a number 
of spectator fields which do not drive the evolution 
of the background but whose inhomogeneities contribute 
to the overall fluctuations of the spatial curvature. Finally, the third 
contribution to the evolution of curvature fluctuations can be 
attributed  to gravitating magnetic fields. 
Up to now various studies addressed the interplay between 
large-scale magnetic fields and the scalar \cite{scalar1,scalar2,scalar3} 
(see also \cite{tsag1,bar1}), vector \cite{vector1,vector2,tensor1}
and tensor \cite{tensor1} modes of the geometry (see \cite{subra} and 
\cite{maxrev} for two recent reviews).  
As far as the scalar modes are concerned, it was assumed 
that large-scale magnetic fields are present prior to recombination 
(but after equality) and the resulting corrections to the Sachs-Wolfe 
plateau have been computed \cite{scalar1,scalar2}. The analysis 
has been also recently extended to cover the Doppler region, i.e.
the first, second and third peaks of the temperature autocorrelations 
\cite{scalar3}.
The main theme of the present paper will be to discuss possible 
(further) numerical constraints arising from the coupled evolution of large-scale 
magnetic fields and curvature perturbations. This analysis, when correctly 
performed, allows to assess the theoretical uncertainty of the 
computational scheme. In fact, 
the results obtained in the case of the 
conventional thermal history can be compared 
with the estimates stemming from less
conventional (but phenomenologically allowed) post-inflationary 
evolutions. 

In the most
simplistic \footnote{Notice that the most simplistic scenario is also 
the one contemplated by the minimal paradigm compatible 
with the three data sets that are used to infer the values 
of the cosmological parameters, i.e.  the $\Lambda$CDM paradigm.
These three data include, in general terms, 
the large-scale structure observations, by the CMB observations 
and by the type Ia supernovae observations. Before plunging into the discussion, it is appropriate to comment on the choice 
of the cosmological parameters that will be employed throughout 
this section. The WMAP data \cite{WMAP1,WMAP2,WMAP3,WMAP4,WMAP5}  have been combined, so far, 
with various sets of data. These data sets include the 2dF 
Galaxy Redshift Survey \cite{2dF}, the combination of Boomerang
and ACBAR data \cite{AC1,AC2}, the combination of CBI and VSA 
data \cite{CB1,CB2}. Furthermore the WMAP 3-year 
data have been also combined with the Hubble Space Telescope 
Key Project (HSTKP) data \cite{HSTKP} as well as with 
the Sloan Digital Sky Survey (SDSS) \cite{SDSS1,SDSS2} data.
Finally, the WMAP 3-year data can be also usefully combined  with 
the weak lensing data \cite{WL1,WL2} and with the observations of  
type Ia supernovae (in particular the data of the Supernova Legacy 
Survey (SNLS) \cite{SNLS} and 
the so-called Supernova "Gold Sample" (SNGS) \cite{SNGS1,SNGS2}). }
scenario  a conventional inflationary phase is followed, after 
a sudden reheating, by a radiation-dominated phase which is replaced, at 
equality by the standard matter-dominated phase. This evolutionary 
history allows to compute the curvature perturbations that are induced, for 
instance, by a single primordial adiabatic mode. It is also possible, in the same 
framework, to include the peculiar effect of large-scale magnetic fields 
whose inhomogeneities may indeed affect the curvature perturbations \cite{scalar1,scalar2,tsag1,bar1}.
The result of this calculation allows the estimate of the Sachs-Wolfe plateau and, ultimately,
the estimate of the Doppler oscillations \cite{scalar3}. 

If the evolutionary history of the background geometry changes in its early phases, also 
the effect of large-scale magnetic fields on the induced curvature perturbations will be different.
The evolution of magnetized curvature perturbations is sensitive, both, to the 
total barotropic index and to the total sound speed, i.e. 
\begin{equation}
w_{\mathrm{t}}(\tau) = \frac{p_{\mathrm{t}}}{\rho_{\mathrm{t}}},\qquad 
c^2_{\mathrm{st}}(\tau) = \frac{p_{\mathrm{t}}'}{\rho_{\mathrm{t}}'},
\label{sounds}
\end{equation}
where the prime denotes a derivation with respect to $\tau$ and the 
subscript t  reminds that the barotropic index and the sound speed 
are the ones pertaining to the {\em total} fluid. 

If the rate of expansion changes in the early 
phases of the thermal history of the Universe (typically prior to big bang nucleosynthesis) also $w_{\mathrm{t}}$ and $c^2_{\mathrm{st}}$ will be 
different and
the curvature perturbations will have a slightly different evolution. 
This qualitative argument will now be corroborated by a detailed calculation and this is one of the themes of the present investigation.
The problem then becomes to estimate the total curvature perturbation that will be sensitive both to the adiabatic and to the magnetized contributions.

The strategy adopted in the present investigation will then be
to scrutinize different thermodynamic histories 
of the Universe and to compute, in each of these cases, the 
resulting curvature perturbations for wavelengths that are still 
larger than the Hubble radius after matter-radiation equality.
It will be shown that possible variations in the evolution of the 
total barotropic index will modify the interplay between the adiabatic and 
the magnetized components of curvature perturbations. This procedure 
will also allow to determine the theoretical error caused by selecting the 
most simplistic thermal history in comparison with its non-minimal
counterparts.

The plan of the present analysis will then be the following. 
In section 2 the essential theoretical tools will be introduced. 
A consistent gauge-invariant description will allow to follow, in one shot,
the evolution of the curvature fluctuations and of the density 
contrasts on comoving orthogonal hypersurfaces.
In section 3 different models for the evolution of the 
barotropic index will be scrutinized and motivated. The resulting (magnetized) curvature perturbations will be computed. In section 4 
a set of bounds on the magnetic field intensity will be derived and compared. Section 5 contains the concluding remarks.

\renewcommand{\theequation}{2.\arabic{equation}}
\setcounter{equation}{0}
\section{Evolution equations}
\label{sec2}
The notations ubiquitously employed in the present script imply that  
the Friedmann-Lema\^itre equations in the spatially flat case (with line 
element given in Eq. (\ref{linel})) are given by:
\begin{eqnarray}
&& {\mathcal H}^2 = \frac{8\pi G}{3} a^2 \rho_{\mathrm{t}},
\label{FL1}\\
&& {\mathcal H}^2 - {\mathcal H}' = 4\pi G a^2 (\rho_{\mathrm{t}} + p_{\mathrm{t}}),
\label{FL2}\\
&& \rho_{\mathrm{t}}' + 3 {\mathcal H} (\rho_{\mathrm{t}} + p_{\mathrm{t}})=0,
\label{FL3}
\end{eqnarray}
where ${\mathcal H} = a'/a$ and the prime denotes, as in Eq. (\ref{sounds}) 
a derivation with respect to $\tau$. Notice that ${\mathcal H} = a H$ 
where $H= \dot{a}/a$ is the Hubble expansion rate and the 
dot denotes a derivation with respect to the cosmic time coordinate $t$ 
(recall that the connection between $t$ and $\tau$ is given by 
the differential relation $d t = a(\tau) d\tau$). 
By virtue of the connection between 
${\mathcal H}$ and $H$, a given wavelength is 
larger than the Hubble radius provided the corresponding comoving 
wave-number $k$ satisfies the condition $k/(a H)\simeq k \tau <1$.
Equations (\ref{sounds}) and (\ref{FL3}) imply
\begin{equation}
c_{\mathrm{st}}^2 = w_{\mathrm{t}} - \frac{w_{\mathrm{t}}'}{ 3 {\mathcal H} (w_{\mathrm{t}} + 1)} = w_{\mathrm{t}} - \frac{1}{3} \frac{d \ln{( w_{\mathrm{t}} + 1)}}{d \ln{a}},
\label{sounds2}
\end{equation}
so that $c_{\mathrm{st}}^2 = w_{\mathrm{t}}$ iff the (total) barotropic 
index is constant in time.
For the purposes of the present investigation it is practical to start with the 
evolution equations of the fluctuations of the geometry expressed in fully gauge-invariant terms. The Hamiltonian 
and momentum constraints can be written as 
\begin{eqnarray}
&&\nabla^2 \Psi - 3 {\mathcal H} ({\mathcal H} \Phi + \Psi') = 4\pi G a^2 (\delta\rho_{\mathrm{t}} + \delta\rho_{\mathrm{B}}),
\label{HAM1}\\
&& \nabla^2 ({\mathcal H} \Phi + \Psi') = - 4\pi G a^2 \,(1 + w_{\mathrm{t}})\, \rho_{\mathrm{t}} \theta_{\mathrm{t}},\qquad \theta_{\mathrm{t}} = \nabla^2 V_{\mathrm{t}},
\label{MOM1}
\end{eqnarray}
where $\delta \rho_{\mathrm{t}}$ is the total (and gauge-invariant) density contrast of the 
fluid sources and $\delta \rho_{\mathrm{B}}(\tau,\vec{x}) = B^2(\vec{x}) /(8\pi a^4)$. The 
fluctuations $\Phi$ and $\Psi$ are the gauge-invariant Bardeen potentials 
\cite{bardeen,bardeen2}
that coincide with the longitudinal fluctuations of the metric in the conformally 
Newtonian gauge. In Eq. (\ref{MOM1}) $\theta_{\mathrm{t}}$ is the 
three-divergence of the total peculiar velocity. At the right-hand 
side of Eq. (\ref{MOM1}) the three-divergence of the Poyniting vector leads 
to a term going as $ \vec{\nabla}\cdot[ \vec{E}\times \vec{B}]/a^4$, where 
$\vec{E} \simeq \vec{J}/\sigma\simeq (\vec{\nabla}\times \vec{B})/\sigma$ 
is the Ohmic electric field. The conductivity $\sigma$ is large during most 
of the thermodynamic history of the Universe, since it is proportional
to the temperature when the temperature is much larger 
than the mass of the correponding species. The contribution of the Ohmic Poynting vector 
will therefore be neglected in comparison with the other components 
of the magnetic energy-momentum tensor.
Introducing the density contrast on comoving orthogonal hypersurfaces, i.e.
\begin{equation}
\epsilon_{\mathrm{m}}(\tau,\vec{x}) = \frac{\delta \rho_{\mathrm{t}} + \delta \rho_{\mathrm{B}}}{\rho_{\mathrm{t}}}  - 3 {\mathcal H} ( 1 + w_{\mathrm{t}}) V_{\mathrm{t}},
\label{epsm1}
\end{equation}
the Hamiltonian constraint (\ref{HAM1}) takes the peculiar form 
\begin{equation}
\nabla^2 \Psi = 4\pi G a^2 \rho_{\mathrm{t}} \epsilon_{\mathrm{m}},
\label{epsm2}
\end{equation}
which is (just formally) analog to the Poisson equation typical of the 
non-relativistic treatment of gravitational inhomogeneities.
Another relevant pair of gauge-invariant quantities is represented by 
\begin{eqnarray}
&&\zeta  = - \Psi - \frac{{\mathcal H}(\delta \rho_{\mathrm{t}} + \delta \rho_{\mathrm{B}})}{\rho_{\mathrm{t}}'},
\label{zeta1}\\
&& {\mathcal R} = - \Psi - \frac{{\mathcal H}({\mathcal H} \Phi + \Psi')}{4\pi G a^2 \rho_{\mathrm{t}}( w + 1)}.
\label{R1}
\end{eqnarray}
The variable $\zeta$ \cite{bardeen2,press} defined in Eq. (\ref{zeta1}), 
if evaluated in the gauge where the spatial curvature 
is uniform, is proportional to the total density contrast (as it can be directly 
checked by using Eq. (\ref{FL3})). The variable ${\mathcal R}$ \cite{press,lyth}, defined in 
Eq. (\ref{R1}), if evaluated on comoving orthogonal hypersurfaces, coincides 
with the fluctuations of the spatial curvature. It is clear that the three 
gauge-invariant variables defined, respectively, in Eqs. (\ref{epsm1}), 
(\ref{zeta1}) and (\ref{R1}) are all related by the Hamiltonian 
constraint. Inserting Eq. (\ref{FL3}) into Eq. (\ref{zeta1}) and taking 
into account Eqs. (\ref{epsm2}), (\ref{R1}) and (\ref{HAM1}) the Hamiltonian 
constraint can be expressed in one of the following two equivalent forms:
\begin{eqnarray}
&& 
\zeta = {\mathcal R} + \frac{\nabla^2 \Psi}{12\pi G a^2 \rho_{\mathrm{t}} ( 1 + 
w_{\mathrm{t}})},
\label{HAM2}\\
&&
\zeta = {\mathcal R} +\frac{\epsilon_{\mathrm{m}}}{3 (1 + w_{\mathrm{t}})}
\label{HAM3},
\end{eqnarray}
where Eq. (\ref{HAM3}) follows from Eq. (\ref{HAM2}) making use of the 
generalized Poisson equation (\ref{epsm2}). 
It is relevant to remark that when the wavelengths are larger than the Hubble 
radius, at a given epoch, 
\begin{equation}
\zeta \simeq {\mathcal R} + {\mathcal O}\biggl(\frac{k^2}{a^2 H^2}\biggr), \qquad 
\epsilon_{\mathrm{m}} \simeq  {\mathcal O}\biggl(\frac{k^2}{a^2 H^2}\biggr).
\label{HAM4}
\end{equation}
In spite of the physical differences between $\zeta$, ${\mathcal R}$ and $\epsilon_{\mathrm{m}}$ the explicit solution of the whole system in terms 
of one of these variables allows to compute the others. 
From the fluctuation of the covariant conservation equation of the 
total plasma the following equation can be easily obtained
\begin{equation}
\delta \rho_{\mathrm{t}}' + 3 {\mathcal H} (\delta \rho_{\mathrm{t}} + \delta 
p_{\mathrm{t}}) - 3 (p_{\mathrm{t}} + \rho_{\mathrm{t}}) \Psi' + (p_{\mathrm{t}}
+\rho_{\mathrm{t}}) \theta_{\mathrm{t}} =0.
\label{cov1}
\end{equation}
Using now Eq. (\ref{zeta1}) inside Eq. (\ref{cov1}) a first-order 
differential equation for $\zeta$ emerges and it is given by:
\begin{equation}
\zeta' = - \frac{{\mathcal H}}{ \rho_{\mathrm{t}} ( 1 + w_{\mathrm{t}})} \delta p_{\mathrm{nad}} + \frac{{\mathcal H}( 3 c_{\mathrm{st}}^2 -1)}{ 3\rho_{\mathrm{t}} ( 1 + w_{\mathrm{t}})}
\delta\rho_{\mathrm{B}} - \frac{\theta_{\mathrm{t}}}{3}.
\label{zeta2}
\end{equation}
The term $\delta p_{\mathrm{nad}}$ in Eq. (\ref{zeta2}) accounts 
for the non-adiabatic pressure inhomogeneities; the following 
decomposition of the total pressure fluctuation
\begin{equation}
\delta p_{\mathrm{t}} = \biggl(\frac{\delta p_{\mathrm{t}}}{\delta \rho_{\mathrm{t}}}\biggr)_{\varsigma} \delta\rho_{\mathrm{t}} +  \biggl(\frac{\delta p_{\mathrm{t}}}{\delta \varsigma}\biggr)_{\rho_{\mathrm{t}}} \delta \varsigma \equiv 
c_{\mathrm{st}}^2 \delta\rho_{\mathrm{t}} + \delta p_{\mathrm{nad}},
\label{deltapnad}
\end{equation}
has been tacitly assumed in Eqs. (\ref{cov1}) and (\ref{zeta2}).
While the first (adiabatic) contribution gives the pressure fluctuation 
produced by the inhomogeneous density fluctuation when the specific 
entropy is constant (i.e. $\delta\varsigma =0$), the second (non-adiabatic) 
contribution arises even if the density is unperturbed (i.e. $\delta \rho_{\mathrm{t}}=0$) but the plasma possesses many different components 
(for instance, in the pre-recombinantion plasma, photons, baryons, neutrinos 
and dark matter).

The evolution  of ${\mathcal R}$ can be obtained, in similar 
terms, from the  equations of $\Phi$ and $\Psi$, i.e.
\begin{eqnarray}
&& \Psi'' + {\mathcal H}(\Phi' + 2 \Psi') + 
( 2 {\mathcal H}' + {\mathcal H}^2) \Phi 
+ \frac{1}{3} \nabla^2 ( \Phi - \Psi) = 4\pi G a^2 (\delta p_{\mathrm{t}} + 
\delta p_{\mathrm{B}}), 
\label{psi1}\\
&& \biggl(\partial_{i}\partial^{j} - \frac{1}{3} \delta_{i}^{j}\nabla^2\biggr) (\Phi - \Psi) = 8\pi G a^2 ( \Pi_{i}^{j} + \overline{\Pi}_{i}^{j}),
\label{psi2}
\end{eqnarray}
where $\Pi_{i}^{j}$ is the anisotropic stress of the fluid and 
\begin{equation}
\overline{\Pi}_{i}^{j} = \frac{1}{4\pi a^4} \biggl( B_{i} B^{j} - \frac{\delta_{i}^{j}}{3}B^2\biggr)
\end{equation}
is the anisotropic stress of the magnetic field.
Inserting Eq. (\ref{R1}) into Eq. (\ref{psi1}) and recalling Eq. (\ref{deltapnad}) 
the evolution equation of ${\mathcal R}$ can be written as 
\begin{equation}
{\mathcal R}' = - \frac{{\mathcal H}}{ \rho_{\mathrm{t}} ( 1 + w_{\mathrm{t}})} \delta p_{\mathrm{nad}} + \frac{{\mathcal H}( 3 c_{\mathrm{st}}^2 -1)}{ 3\rho_{\mathrm{t}} ( 1 + w_{\mathrm{t}})}
\delta\rho_{\mathrm{B}} + \frac{{\mathcal H}}{12 \pi G a^2 (p_{\mathrm{t}} + \rho_{\mathrm{t}})} \nabla^2 (\Phi - \Psi) - \frac{{\mathcal H} c_{\mathrm{st}}^2}{4\pi G a^2 (p_{\mathrm{t}} + \rho_{\mathrm{t}})} \nabla^2 \Psi.
\label{R2}
\end{equation}
By now subtracting Eq. (\ref{R2}) from Eq. (\ref{zeta2}) and by using Eqs. 
(\ref{HAM3}) and (\ref{HAM4}), the appropriate equation of 
 $\epsilon_{\mathrm{m}}$ can be obtained after simple algebra:
\begin{equation}
\epsilon_{\mathrm{m}}' - 3 {\mathcal H} w_{\mathrm{t}}
 \epsilon_{\mathrm{m}} = - ( 1 + w_{\mathrm{t}}) \theta_{\mathrm{t}} - 
 3 {\mathcal H}( 1 + w_{\mathrm{t}}) \Pi_{\mathrm{t}}
 \end{equation}
 where the notation 
 \begin{equation}
 \partial_{i}\partial^{j}\Pi_{j}^{i} +  \partial_{i}\partial^{j}\overline{\Pi}_{j}^{i}
 = (p_{\mathrm{t}} + \rho_{\mathrm{t}}) \nabla^2 \Pi_{\mathrm{t}},
 \label{Piij}
\end{equation}
has been used. If there are collisionless particles in the plasma 
(like neutrinos, after weak interactions have fallen out of thermal equilibrium) we will have that 
\begin{equation}
(p_{\mathrm{t}} + \rho_{\mathrm{t}}) \nabla^2 \Pi_{\mathrm{t}} = (p_{\nu} + \rho_{\nu} ) \nabla^2 \Pi_{\nu} + (p_{\gamma} + \rho_{\gamma}) \nabla^2 \Pi_{\mathrm{B}}.
\label{PPB}
\end{equation}
Concerning Eq. (\ref{PPB}) two comments are in order. We referred 
the magnetic anisotropic stress to the photon background. This 
is natural since, for typical length-scales larger than the magnetic diffusivity 
scale, both quantities redshift at the same rate with the expansion 
of the Universe. The second comment involves the relevance 
of the anisotropic stress. As discussed in relation with the estimate of 
the CMB autocorrelation induced by magnetic fields, the magnetic anisotropic 
stress is of upmost importance at intermediate scales. In particular the interplay with the neutrino anisotropic stress leads to the correct 
initial conditions for the magnetized CMB anisotropies. However, in the present paper we will be mostly concerned with the largest scales and 
we will try to assess what could be the influence, on those scales, of slightly 
different thermal histories of the Universe. For typical wavelengths 
larger than the Hubble radius at recombination the anisotropic stress 
can be neglected in the first approximation. If needed, however, it can be 
included with the standard iterative procedure \cite{hu1,hu2} (see also \cite{kodama}) 
where, to lowest order, the solution is the one where $\Phi \simeq \Psi$.
\renewcommand{\theequation}{3.\arabic{equation}}
\setcounter{equation}{0}
\section{Different thermal histories}
\label{sec3}
By looking at Eq. (\ref{zeta2}) it is clear that, to leading order 
in $k^2 \tau^2$, the evolution of $\zeta$ (and of ${\mathcal R}$) is determined
by three independent pieces of information:
\begin{itemize}
\item{} the presence (or absence) of non-adiabatic pressure fluctuations;
\item{} presence (or absence) of large-scale magnetic fields;
\item{} the specific time dependence of $w_{\mathrm{t}}$ and $c_{\mathrm{st}}^2$
\end{itemize}
The simplest possible situation compatible with 
the presence of super-Hubble magnetic fields arises when the Universe 
becomes suddenly dominated by radiation at the end of inflation. 
This is, incidentally, also the underlying assumption in the standard 
$\Lambda$CDM scenario.
The Universe will then become dominated by dusty matter at a 
redshift\footnote{When not otherwise stated it will be assumed that 
$h_{0}^2 
\Omega_{\mathrm{M}0} \simeq 0.1326$, where $\Omega_{\mathrm{M}0}$ 
denotes the present fraction in dusty matter. This value for the 
total critical fraction of matter emerges, in the context of the $\Lambda$CDM paradigm, when the WMAP data \cite{WMAP1,WMAP2} are combined with all the remaining cosmological data sets of different origin (excluding weak lensing data).}
  $z_{\mathrm{eq}} \simeq 3200$.
If inflation was driven by a single inflaton and if other spectator fields 
with scale-invariant spectrum were absent, then it is also rather 
plausible to enforce the condition
$\delta p_{\mathrm{nad}}=0$.  
In this situation, the exact solution of Eqs. (\ref{FL1}), (\ref{FL2}) and (\ref{FL3}) implies that 
\begin{equation}
w_{\mathrm{t}}(\alpha)= \frac{1}{3(\alpha + 1)},\qquad 
c_{\mathrm{st}}^2(\alpha) = \frac{4}{3 (3 \alpha + 4)},
\label{speeds3}
\end{equation}
where 
\begin{eqnarray}
&& \alpha(\tau) = \frac{a}{a_{\mathrm{eq}}} = x^2 + 2 x,\qquad x= \frac{\tau}{\tau_{1}},\frac{a_{0}}{a_{\mathrm{eq}}} =\frac{h_{0}^2 
\Omega_{\mathrm{M}0}}{h_{0}^2 
\Omega_{\mathrm{R}0}},
\nonumber\\
&& \frac{a_{0}}{a_{\mathrm{eq}}} =\frac{h_{0}^2 
\Omega_{\mathrm{M}0}}{h_{0}^2 
\Omega_{\mathrm{R}0}}, \qquad \tau_{\mathrm{eq}} = (\sqrt{2}-1)\tau_{\mathrm{eq}}.
\label{alpha1}
\end{eqnarray}
Using Eqs. (\ref{speeds3}) and (\ref{alpha1}) inside Eq. (\ref{zeta2}) 
and changing the variable from $\tau$ to $\alpha$ the following expression
can be obtained:
\begin{equation}
\frac{d \zeta}{d\alpha} = - \frac{ 3 R_{\gamma} \Omega_{\mathrm{B}}}{(3 \alpha + 4)^2},\qquad R_{\gamma} = \frac{\rho_{\gamma}(\tau)}{\rho_{\mathrm{R}}(\tau)}
\qquad 
\Omega_{\mathrm{B}}(\tau,\vec{x})= \frac{\delta\rho_{\mathrm{B}}(\tau,\vec{x})}{\rho_{\gamma}(\tau)}.
\label{RM1}
\end{equation}
In Eq. (\ref{RM1}) the magnetic energy density has been expressed in units 
of the photon energy density and $R_{\gamma}$ denotes the photon fraction 
in the post-inflationary radiation background. After neutrinos 
decouple for temperatures of the order of the MeV the photon fraction 
will be related with the neutrino fraction as $R_{\gamma} = 1 - R_{\nu}$ 
where $R_{\nu} = 0.405$ (for three neutrino families).
Direct integration of Eq. (\ref{RM1}) gives the sought result, namely
\begin{equation}
\zeta(k,\tau)= \zeta_{*}(k) - \frac{3 R_{\gamma} \Omega_{\mathrm{B}(k)} \alpha(\tau) }{4 ( 3\alpha(\tau) + 4)}.
\label{RM2}
\end{equation}
The constant $\zeta_{*}(k)$ stands for the adiabatic mode 
produced (for instance during a phase of conventional inflation). After 
matter-radiation equality (but before recombination) to leading order 
in $k^2 \tau^2$ we will have, from Eq. (\ref{HAM2}), that 
\begin{equation}
{\mathcal R}(k,\tau_{\mathrm{rec}}) \simeq \zeta_{*}(k) - \frac{R_{\gamma} \Omega_{\mathrm{B}}(k)}{4}
 \biggl(\frac{ 3 \alpha_{\mathrm{rec}}}{3 \alpha_{\mathrm{rec}} + 4}\biggr) + {\mathcal O}(k^2 \tau_{\mathrm{rec}}^2), \qquad \alpha_{\mathrm{rec}} \simeq \frac{z_{\mathrm{eq}} +1}{z_{\mathrm{rec}}+1} = 3.07 \,\biggl(\frac{h_{0}^2 \Omega_{\mathrm{M}0}}{0.134}\biggr).
\label{RM3}
\end{equation}
This result can be easily interpreted from the physical point of view. When the rate of expansion 
increases from radiation to matter the barotropic index (and the sound speed) are both decreasing 
from $1/3$ to $0$. Thus the overall effect on the source term for the evolution of curvature perturbations 
implies that the magnetic contribution tends to cancel the contribution of the adiabatic mode.

In the context of this reference scenario when the Universe only passes, after 
inflation, from radiation to matter also $\epsilon_{\mathrm{m}}$ and $\theta_{\mathrm{t}}$ can be computed, respectively, from Eqs. (\ref{epsm2}) and (\ref{MOM1}). It suffices to extract $\Psi$ from $\zeta$ and then use 
the aforementioned constraints. Going to Fourier space result of this manipulation will be 
\begin{eqnarray}
&& \epsilon_{\mathrm{m}}(k,\tau) = \frac{k^2 \tau_{1}^2}{12 \alpha \sqrt{\alpha+1}} \biggl[ \zeta_{*}(k) {\mathcal W}_{1}(\alpha)- \frac{3}{4} R_{\gamma} \Omega_{\mathrm{B}}(k) {\mathcal W}_{2}(\alpha)\biggr],
\label{h1eps}\\
&& \theta_{\mathrm{t}}(k,\tau) = -\frac{k^2\tau_1}{2}\biggl\{\biggl[\frac{\alpha}{\sqrt{\alpha+1}} - \frac{{\mathcal W}_{1}(\alpha)}{2 \alpha^2}\biggr] 
\zeta_{*}(k)+ 
\frac{3}{8} R_{\gamma} \Omega_{\mathrm{B}}(k)\biggl[ 
\frac{{\mathcal W}_{2}(\alpha)}{\alpha^2} - \frac{2 \alpha^2}{(3 \alpha + 4) \sqrt{\alpha + 1}}\biggr]\biggr\},
\label{h1th}
\end{eqnarray}
where 
\begin{eqnarray}
&& {\mathcal W}_{1}(\alpha) = \frac{2}{15 \sqrt{\alpha + 1}} \{ 16 [\sqrt{\alpha} -1] + \alpha[ \alpha ( 9 \alpha + 2) -8]\},
\nonumber\\
&& {\mathcal W}_{2}(\alpha) = \frac{2}{5 \sqrt{\alpha + 1}} \{ 16 [ 1 - \sqrt{\alpha + 1}] + \alpha [ 8 + \alpha (\alpha -2)]\}.
\label{Wh1}
\end{eqnarray}

Thanks to the success of big-bang nucleosynthesis, it is rather plausible
to imagine that the Universe was already dominated by radiation for temperatures as large and few MeV. However, prior to that time there 
are no direct probes of the expansion rate of the Universe. In a somehow 
indirect way this could be achieved through the study of the cosmic 
background of gravitational waves. At the moment, for instance, there is 
no compelling evidence of the fact that, after inflation, the Universe became suddenly dominated by radiation. It is, on the contrary, rather reasonable 
that, before settling on a radiation-dominated stage of expansion 
the Universe passed through a phase where, for instance, the rate 
of expansion was smaller than the Hubble rate during radiation. This 
occurrence is realized when the total barotropic index of the sources driving 
the geometry satisfies $ 1/3< w_{\mathrm{t}} \leq 1$. In this case the sources 
are said to be stiffer than radiation \cite{stiffer}. The stiffest equation of state we can imagine is the one where the barotropic index is $1$ which implies that 
the sound speed coincides with the speed of light. The case of speed of sound equal to speed of light 
is the one contemplated by the so-called Zeldovich model (see \cite{grishchuk} and references 
therein).  A similar kind of evolution arises in quintessential inflationary models 
where the inflaton and the quintessence field are unified in a single degree of freedom. 
In this case (as in the more general case of stiffer phases \cite{stiffer}) the spectrum of relic gravitons 
presents a sharp increase \cite{quint1} and a peak \cite{quint2} which
are potentially accessible to direct observations.  
In the context of quintessential 
inflation the stiff epoch is effectively dominated by the kinetic energy of the inflaton. Stiff 
phases also arise in brane-world scenarios \cite{sa1} where the calculation of the graviton spectrum 
mirrors the four-dimensional case \cite{quint1,quint2}.

The transition from a stiff epoch to a radiation-dominated 
phase of expansion can be understood in terms of the following exact solution:
\begin{equation}
\alpha = \frac{a}{a_{\mathrm{r}}} = \sqrt{y^2 + 2 y}, \qquad y = \frac{\tau}{\tau_{1}},\qquad \tau_{1} = \frac{2}{\lambda H_{*}},
\label{stiff1}
\end{equation}
where
\begin{equation}
a_{\mathrm{r}} = a_{*} \sqrt{\frac{\rho_{\mathrm{s}*}}{\rho_{\mathrm{R}*}}} \simeq \frac{M_{\mathrm{P}}}{H_{*}}, \qquad \lambda= a_{*} \frac{H_{*}}{M_{\mathrm{P}}}
\label{stiff2}
\end{equation}
Equations (\ref{stiff1}) and (\ref{stiff2}) are solutions of Eqs. (\ref{FL1}), (\ref{FL2}) and (\ref{FL3}) when 
the sources of the geometry are given in terms of a mixture of radiation and of a stiff fluid with $p_{\mathrm{s}} = \rho_{\mathrm{s}}$. 
Equations (\ref{stiff1}) and (\ref{stiff2}) parametrize the physical situation 
where, at $\tau_{*}$ some (small) amount of radiation is present 
in comparison with the stiff contribution (i.e. $\rho_{\mathrm{R}*}\ll 
\rho_{\mathrm{s}*}$). In the context of quintessential inflationary models 
\cite{pee} the initial amount of radiation comes indeed from 
quantum fluctuations \cite{ford} and, consequently, $\rho_{\mathrm{R}*}\simeq H_{*}^4$. This occurence implies, in turn, that the hierarchy 
between the initial scale factor at $\tau_{*}$ and the scale factor at the 
onset of radiation (i.e. $a_{\mathrm{r}}$ is determined by $M_{\mathrm{P}}/H_{*}$). Since $H_{*} <10^{-6} M_{\mathrm{P}}$, the duration of the stiff
phase is correctly bounded because $H_{\mathrm{r}} > H_{\mathrm{bbn}}$.
According to Eq. (\ref{sounds2}), the 
barotropic index and the sound speed are:
\begin{equation}
w_{\mathrm{t}} = \frac{1}{3} \frac{\alpha^2 + 3}{\alpha^2 +1}, \qquad c_{\mathrm{st}}^2 = \frac{2 \alpha^2 + 9}{3( 2 \alpha^2 + 3)}.
\label{stiff3}
\end{equation}
Using therefore Eq. (\ref{stiff3}) inside Eq. (\ref{zeta2}) and changing variable from $\tau$ to $\alpha$:
\begin{equation}
\frac{d\zeta}{d\alpha} = \frac{3 R_{\gamma} \Omega_{\mathrm{B}} \alpha}{(2 \alpha^2 + 3)^2}.
\label{stiff4}
\end{equation}
Integrating Eq. (\ref{stiff4}) between $\alpha=0$ and $\alpha_{\mathrm{f}}$ we do get 
\begin{equation}
\zeta(k,\tau) = \zeta_{*}(k) + \frac{R_{\gamma} \Omega_{\mathrm{B}(k)} \, \alpha_{\mathrm{f}}^2}{2 (2 \alpha_{\mathrm{f}}^2 + 3)},
\label{stiff5}
\end{equation}
where $\alpha_{\mathrm{f}}\gg 1 $ and it is defined deep in the radiation epoch. This 
result implies that the magnetic contribution enhances (rather than canceling) the adiabatic term 
$\zeta_{*}(k)$ of inflationary origin. 

As in the case of the radiation-matter transition the total density contrast and the total peculiar velocity can be computed. 
In Fourier space the result is:
\begin{eqnarray}
&&\epsilon_{\mathrm{m}}(k,\tau) 
= - \frac{k^2 \tau_{*}^2}{6 \sqrt{\alpha^2 +1 }} [ 2 
\zeta_{*}(k) {\mathcal W}_{3}(\alpha) + R_{\gamma} \Omega_{\mathrm{B}}(k) {\mathcal W}_{4}(\alpha)],
\label{h2eps}\\
&& \theta_{\mathrm{t}}(k,\tau) = - k^2 \tau_{*} \biggl\{ \biggl[ \frac{\alpha^2}{\sqrt{\alpha^2 +1}} - \frac{2 {\mathcal W}_{3}(\alpha)}{\alpha^2}\biggr] \zeta_{*}(k) 
\nonumber\\
&&+ \frac{R_{\gamma} \Omega_{\mathrm{B}}(k)}{2} \biggl[
\frac{\alpha^4}{( 2 \alpha^2 + 3)\sqrt{\alpha^2+1}} - \frac{2}{\alpha^2} {\mathcal W}_{4}(\alpha)\biggr]\biggr\},
\label{h2th}
\end{eqnarray}
where 
\begin{eqnarray}
&& {\mathcal W}_{3}(\alpha) = \frac{4}{3} [ (\alpha^2 +1)^{3/2} -1] - \frac{2 \alpha^2}{\sqrt{\alpha^2 +1 }},
\nonumber\\
&& {\mathcal W}_{3}(\alpha) = \frac{2}{3} [ (\alpha^2 +1 )^{3/2} + 8] - 
\frac{2 (2\alpha^2 + 3)}{\sqrt{\alpha^2 + 1}}.
\end{eqnarray}

This result can be generalized when the stiff phase is parametrized in terms of a source $p_{\mathrm{s}} = \delta \rho_{\mathrm{s}}$ with $1 < \delta < 1/3$. Repeating the same steps outlined above we will have that 
the dependence of the barotropic index and of the sound speed upon the scale factor are given, in this case,  by:
\begin{equation}
w_{\mathrm{t}} = \frac{1}{3} \biggl(\frac{3 \delta + \alpha^{3 \delta -1}}{1 + \alpha^{3 \delta -1}}\biggr), 
\qquad c_{\mathrm{st}}^2 = \frac{ 9 \delta (\delta + 1) + 4 \alpha^{3 \delta -1}}{3 [ 3 (\delta + 1) + 4 \alpha^{3\delta -1}]}.
\label{stiff6}
\end{equation}
The equation for $\zeta$ and its related solution can the be written as 
\begin{eqnarray}
&& \frac{d\zeta}{d\alpha} = 3 R_{\gamma} \Omega_{\mathrm{B}} \frac{(\delta + 1) (3 \delta -1)\, \alpha^{3\delta -2}}{[3 (\delta + 1) + 4 \alpha^{ 3 \delta -1}]^2},
\label{stiff7}\\
&& \zeta(k,\tau)= \zeta_{*}(k) + R_{\gamma} \Omega_{\mathrm{B}}(k) \frac{\alpha_{\mathrm{f}}^{3 \delta-1}}{4 \alpha_{\mathrm{f}}^{ 3\delta -1} + 3 (\delta + 1)}.
\label{stiff8}
\end{eqnarray}
As in the other two cases $\epsilon_{\mathrm{m}}$ and $\theta_{\mathrm{t}}$ 
can be easily computed:
\begin{eqnarray}
&& \epsilon_{\mathrm{m}}(k,\tau) = \frac{k^2 \tau_{*}^2}{9 \delta^2 -1} 
\frac{\alpha^{3(\delta -1)/2}}{\sqrt{ 1 + \alpha^{3\delta -1}}}[ 
{\mathcal W}_{5}(\alpha,\delta) \zeta_{*}(\vec{k}) + R_{\gamma} 
{\mathcal W}_{6}(\alpha,\delta) \Omega_{\mathrm{B}}(k)],
\label{h3eps}\\
&& \theta_{\mathrm{t}}(k,\tau) = \frac{k^2 \tau_{*}}{\sqrt{3 \delta + 1}} 
\frac{\alpha^{(3\delta + 1)/2}}{\sqrt{\alpha^{3\delta -1} +1}}\biggl\{ 
\zeta_{*}(k) \biggl[ 1 - \frac{\sqrt{\alpha^{3 \delta -1} + 1}}{2 (3 \delta -1)} 
\frac{{\mathcal W}_{5}(\alpha,\delta)}{\alpha^{(3\delta + 5)/2}}\biggr]
\nonumber\\
&& R_{\gamma} \Omega_{\mathrm{B}}(\vec{k}) \biggl[ \frac{\alpha^{3 \delta -1}}{4 \alpha^{3 \delta -1} + 3 (\delta + 1)} - \frac{\sqrt{\alpha^{3 \delta -1} +1}}{\alpha^{(3 \delta + 5)/2}} \frac{{\mathcal W}_{6} (\alpha,\delta)}{2 ( 3 \delta -1)}\biggr] \biggl\},
\label{h3th}
\end{eqnarray}
where
\begin{eqnarray}
&&{\mathcal W}_{5}(\alpha,\delta)= \int_{0}^{s(\alpha)} \frac{ 4 s + 3 (\delta+1)}{(s+ 1)^{3/2}} s^{\frac{7 - 3\delta}{2 (3 \delta -1)}} ds,
\nonumber\\
&& {\mathcal W}_{6}(\alpha,\delta)= \int_{0}^{s(\alpha)}
\frac{s^{\frac{5 + 3 \delta}{2 (3 \delta -1)}}}{(s + 1)^{3/2}} ds.
\label{inth3}
\end{eqnarray}
The upper limit of integration in Eqs. (\ref{inth3}) 
corresponds to $s(\alpha) = \alpha^{3\delta -1}$. For each value of $\delta$ 
the above integrals can be evaluated in explicit terms.

Suppose now to investigate a three stage model where the Universe expanded, initially, at a rate that was 
slower than radiation then got trapped in a radiation phase which turned subsequently into an epoch dominated 
by dusty matter. In this case we will have that the barotropic index and the sound speed 
are given, respectively, by:
\begin{eqnarray}
&& w_{\mathrm{t}} = \frac{1}{3} \frac{ 3 + \alpha^2}{1 + \alpha^2 + b \alpha^3},\qquad b = \frac{\rho_{\mathrm{M}*}}{\rho_{\mathrm{s}*}} \biggl(\frac{\rho_{\mathrm{s}*}}{\rho_{\mathrm{R}*}}\biggr)^{3/2},
\label{stiff9}\\
&& c_{\mathrm{st}}^2 = \frac{ 18 + 4\alpha^2}{18 + 12 \alpha^2 + 9 b \alpha^3}.
\label{stiff10}
\end{eqnarray}
In this case the evolution equation of curvature perturbations is given by 
\begin{equation}
\frac{d\zeta}{d\alpha} = - 3 R_{\gamma} \Omega_{\mathrm{B}} \frac{\alpha (\alpha^3 b - 4)}{(3\alpha^3 b + 4 \alpha^2 + 6)^2}.
\label{stiff11}
\end{equation}
The (approximate) solution of Eq. (\ref{stiff11}) in the matter-dominated epoch reads:
\begin{equation}
\zeta(k,\tau) \simeq \zeta_{*}(k) + \frac{R_{\gamma} \Omega_{\mathrm{B}}(k)}{3 \alpha_{\mathrm{rec}} + 4}.
\label{stiff12}
\end{equation}
The result of  Eq. (\ref{stiff12}) 
shows an extra suppression of a factor $\alpha_{\mathrm{rec}}\simeq 1/3$ in comparison 
with Eq. (\ref{RM3}). This suppression can be easily understood. The evolution described by Eqs. (\ref{stiff9}) 
and (\ref{stiff10}) implies that, right after inflation, the Universe expands slower than radiation.
When radiation kicks in, the curvature perturbations are enhanced. Later on, when dusty matter 
becomes dominant (around matter-radiation equality) the magnetized contribution 
tends, again, to cancel the pre-existing adiabatic mode. The net result of the initial increase 
and of the subsequent decrease of $\zeta$ is given in Eq. (\ref{stiff12}) and it is a bit different from the result 
of Eq. (\ref{RM3}) where the intermediate stiff phase was absent. 
The following conclusions can then be drawn:
\begin{itemize}
\item{} if the Universe is dominated by radiation from the end of inflation (as it happens in the case 
of the standard evolution) the magnetized contribution and the adiabatic mode have opposite sign;
\item{} if, prior to the electroweak epoch, the thermodynamic history of the Universe 
deviates from a radiation epoch the magnetized contribution is more suppressed in 
comparison with the standard case.
\end{itemize}
In the framework of conventional inflationary models this type of deviation from the 
standard scenario is pretty general. By general we mean that, within the current bounds 
on the expansion of the Universe arising from big-bang nucleosynthesis we can just 
imagine drastic deviation from the radiation-dominated evolution between the end of inflation 
and, say, neutrino decoupling. In fact, deviation from a radiation-dominated evolution 
after (or right before) big bang nucleosyntheis will necessarily jeopardize the 
production of the abundances of the light elements. 
 
Up to now it has been assumed that the fluids of the primeval plasma do not exchange 
energy. Let us  now address this interesting situation that could arise, for instance, in the 
course of reheating. Let us therefore imagine that a matter fluid (which can model the coherent 
oscillations of the inflaton) decays into massless particles. This dynamics can be 
parametrized by the following system \footnote{Here we passed from the conformal to the cosmic 
time parametrization. See the comments after Eqs. (\ref{FL1}), (\ref{FL2}) and (\ref{FL3}).}:
\begin{eqnarray}
&& \dot{H} = - 4\pi G( \frac{4}{3} \rho_{\rm r} + \rho_{\rm m}),
\label{decb1}\\
&& \dot{\rho}_{\rm m} = - 3 H \rho_{\rm m} - \Gamma \rho_{\rm m},
\label{decb2}\\
&& \dot{\rho}_{\rm r} = - 4 H \rho_{\rm r} + \Gamma \rho_{\rm m}.
\label{decb3}
\end{eqnarray}
The Universe, initially dominated (right after inflation) by dusty 
matter, becomes rapidly dominated by radiation at a rate that is controlled by $\overline{\Gamma}$.
It should be immediately noticed that, by summing up Eqs. (\ref{decb2}) and (\ref{decb3}) the total 
energy density $\rho_{\mathrm{t}}= \rho_{\mathrm{r}} + \rho_{\mathrm{m}}$ is covariantly conserved, i.e. 
$\dot{\rho}_{\mathrm{t}} + 3 H (\rho_{\mathrm{t}} + p_{\mathrm{t}})=0$.  Equations (\ref{decb2}) and (\ref{decb3}) 
can be approximately solved:
\begin{equation}
\rho_{\mathrm{m}}(t) = \rho_{\mathrm{r}}(t_{\mathrm{i}}) \biggl(\frac{a_{\mathrm{i}}}{a}\biggr)^3 e^{- \overline{\Gamma}(t - t_{\mathrm{i}})}, \qquad \rho_{\mathrm{r}}(t) \simeq\rho_{\mathrm{r}}(t_{\mathrm{d}}) \biggl(\frac{a_{\mathrm{d}}}{a}\biggr)^{4},
\label{decb4}
\end{equation}
where $t_{\mathrm{d}} \simeq \overline{\Gamma}^{-1}$. As a consequence of the relations of Eq. (\ref{decb4}) 
we can also say that 
\begin{equation}
\frac{\rho_{\rm m}}{\rho_{\rm r}} \simeq \frac{H_{\mathrm{i}}^2}{H_{\mathrm{d}}^2} \biggl(\frac{a_{\mathrm{i}}}{a}\biggr)^3 \biggl( \frac{a}{a_{\mathrm{d}}}\biggr)^4 e^{- \overline{\Gamma}(t - t_{\mathrm{i}})}
\simeq (t \overline{\Gamma})^{1/2} e^{- \overline{\Gamma}(t - t_{\mathrm{i}})}
\label{decb5}
\end{equation}
In this situation the total barotropic index 
and the total sound speed are slightly modified thanks to the physical differences of the system. In particular we will have that
\begin{equation}
c_{\mathrm{st}}^2 = \frac{4}{3} \frac{\rho_{\mathrm{r}}}{4 \rho_{\mathrm{r}} + 3 \rho_{\mathrm{m}}} - 
\frac{\overline{\Gamma}}{3H} \frac{\rho_{\mathrm{m}}}{4 \rho_{\mathrm{r}} + 3 \rho_{\mathrm{m}}}.
\label{decb6}
\end{equation}
From Eq. (\ref{decb6}) is clear that $c_{\mathrm{st}}^2 \to 1/3$ when $\overline{\Gamma}\gg H$.

The evolution equations of curvature perturbations to be solved read:\begin{eqnarray}
&& \dot{\zeta} = - \frac{H}{p_{\mathrm{t}} + \rho_{\mathrm{t}}} \delta p_{\mathrm{nad}} + \frac{H}{p_{\mathrm{t}} + \rho_{\mathrm{t}}} 
\biggl(c_{\mathrm{st}}^2 - \frac{1}{3}\biggr) \delta\rho_{\mathrm{B}},
\label{deca1}\\
&& \dot{\zeta}_{\rm m} = \frac{\dot{g}}{g} \zeta_{\rm m} - \overline{\Gamma} g
\frac{\dot{H}}{H^2} \zeta,
\label{deca2}
\end{eqnarray}
where $g = H/(\overline{\Gamma} + 3 H)$.
Let us now concentrate, as discussed above, on the conventional situation where the non-adiabatic modes 
are totally absent. In this case, initially, $\delta p_{\mathrm{nad}} \propto (\zeta_{\mathrm{r}} - \zeta_{\mathrm{m}}) \simeq 0$ and $\zeta_{\mathrm{r}} = \zeta_{\mathrm{m}} \simeq \zeta_{*}$. 
Using Eq. (\ref{decb6}) inside Eq. (\ref{deca1}) we will simply have that 
\begin{equation}
\dot{\zeta} = - \frac{H \delta\rho_{\mathrm{B}} \rho_{\mathrm{m}}}{g ( 4 \rho_{\mathrm{r}} + 3 \rho_{\mathrm{m}})^2}
\label{deca3}
\end{equation}
Recalling that, after $t_{\mathrm{d}}$, $\overline{\Gamma}\gg H$ and $\rho_{\mathrm{m}}$ is 
exponentially suppressed, Eq. (\ref{deca3}) can be written as 
\begin{equation}
\frac{ d \zeta}{d x} = - \frac{R_{\gamma} \Omega_{\mathrm{B}}}{16} \sqrt{x} e^{-x},\qquad x = \overline{\Gamma} t.
\label{deca4}
\end{equation}
Integrating once Eq. (\ref{deca4}) the result will be 
\begin{equation}
\zeta(k,y) = \zeta_{*}(k)-  \frac{R_{\gamma} \Omega_{\mathrm{B}}(k)}{16}  \int_{1}^{y} \sqrt{x} e^{- x}\,\, d\,x. 
\end{equation}
The integral mentioned above can be done introducing the error function, i.e. 
\begin{equation}
 \int_{1}^{y} \sqrt{x} e^{- x}\,\, d\,x=
\frac{1}{e} - \frac{{\sqrt{y}}}{e^y} - \frac{{\sqrt{\pi }}\,{\rm Erf}(1)}{2} + \frac{{\sqrt{\pi }}\,{\rm Erf}({\sqrt{y}})}{2}.
\end{equation}
Taking the limit $y\to \infty$, we then have 
\begin{equation}
\zeta(k,t) \to \zeta_{*}(k)-  \frac{R_{\gamma} \Omega_{\mathrm{B}}(k)}{16} \biggl(\frac{2 + e\,{\sqrt{\pi }} - e\,{\sqrt{\pi }}\,{\rm Erf}(1)}{2\,e}\biggr).
\end{equation}
that is, numerically, 
\begin{equation}
\zeta(k,t) \sim \zeta_{*}(k)- 0.079\, R_{\gamma} \Omega_{\mathrm{B}}(k)
\sim \zeta_{*}(k) - \frac{R_{\gamma}\Omega_{\mathrm{B}}(k)}{12}.
\label{deca5}
\end{equation}

\renewcommand{\theequation}{4.\arabic{equation}}
\setcounter{equation}{0}
\section{Bounds on different thermal histories}
\label{sec4}
The two-point correlation function of curvature perturbations 
will receive contributions, in general, from the adiabatic mode, from one (or more) non-adiabatic modes,  and from the magnetic field. As argued 
in the previous sections, the simplest situation is the one where 
only the adiabatic mode is present together with the magnetized 
contribution that we ought to constrain. This choice is motivated by the 
experimental evidence that the Doppler peak structure of the temperature 
autocorrelations strongly suggests that, after equality, the large-scale 
curvature perturbations were, predominantly adiabatic \cite{WMAP1,WMAP2,WMAP3,WMAP4,WMAP5}. 

Consider therefore, the curvature perturbations present at recombination 
in the framework of the different thermal histories discussed in the previous section. The curvature perturbations can be written, in Fourier space, as
\footnote{The quantity $T_{h}(\alpha_{\mathrm{rec}}, h_{0}^2 \Omega_{\mathrm{M}0})$ is the generalized transfer function that may change 
depending upon the specific thermal history labeled by the subscript 
$h$.}
\begin{equation}
\zeta_{h}(k,\alpha_{\mathrm{rec}}) = \zeta_{*}(k)  + 
T_{h}(\alpha_{\mathrm{rec}}, h_{0}^2 \Omega_{\mathrm{M}0}) \Omega_{\mathrm{B}}(k),
\label{b1}
\end{equation}
where the subscript $h$ refers to each different thermal history and where 
$\zeta_{*}(k)$ represents the adiabatic contribution 
normalized to the large-scale value of the (ordinary) Sachs-Wolfe 
contribution to the temperature autocorrelations.
The total (scalar) power spectrum ${\mathcal P}_{h}(k)$, i.e. the Fourier transform of the two-point 
function computed at different spatial locations but at the same time is 
defined as
\begin{equation}
\langle \zeta_{h}(\tau, \vec{x}) \zeta_{h}(\tau, \vec{y}) \rangle =
\int d \ln{k} {\mathcal P}_{h}(k) \frac{\sin{k r}}{k r}, \qquad r = |\vec{x} - \vec{y}|.
\label{b2}
\end{equation}
From Eq. (\ref{b1}) and within the conventions summarized by Eq. (\ref{b2}) 
we then have that
\begin{equation}
\langle \zeta_{h}(\vec{k},\tau) \zeta_{h}(\vec{p},\tau) = 
\frac{2\pi^2}{k^2} {\mathcal P}_{h}(k) \delta^{(3)}(\vec{k}+ \vec{p})
\label{b3}
\end{equation}
where
\begin{equation}
{\mathcal P}_{h}(k) = {\mathcal P}_{\zeta}(k) + T_{h}^2(\alpha_{\mathrm{rec}}, 
h_{0}^2 \Omega_{\mathrm{M}0}) {\mathcal P}_{\Omega}(k) + 
2 T_{h}(\alpha_{\mathrm{rec}}, h_{0}^2 \Omega_{\mathrm{M}0}) 
\sqrt{{\mathcal P}_{\zeta}(k)} \sqrt{{\mathcal P}_{\Omega}(k)} \cos{\gamma},
\label{b4}
\end{equation}
where $\gamma$ is the correlation angle between the adiabatic 
mode and the magnetized contribution.
In Eq. (\ref{b4}) ${\mathcal P}_{\zeta}$ is the power spectrum of the 
adiabatic contribution defined as 
\begin{equation}
{\mathcal P}_{\zeta} = {\mathcal A}_{\zeta} \biggl(\frac{k}{k_{\mathrm{p}}}\biggr)^{n_{\zeta} -1}, \qquad {\mathcal A}_{\zeta} =\frac{2\times 10^{4}}{9 T_{\mathrm{cmb}}^2} A(k_{\mathrm{p}}) = 2.95 \times 10^{-9} A(k_{\mathrm{p}}).
\label{b5}
\end{equation}
In Eq. (\ref{b5}) $n_{\zeta}$ is the spectral index of the adiabatic mode 
and ${\mathcal A}_{\zeta}$ is the amplitude referred to the pivot scale 
$k_{\mathrm{p}}$. Following the WMAP conventions the value 
of the pivot scale will be fixed as $k_{\mathrm{p}} = 0.002 \, {\mathrm{Mpc}}^{-1}$. The numerical factor appearing in the expression of ${\mathcal A}_{\zeta}$ has been obtained by taking consistently $T_{\mathrm{cmb}} = 2.725 
\times 10^{6}$ (expressed in units of $\mu K$). In the absence of any other contributions
the WMAP 3-year data imply\footnote{In this analysis 
the values of the cosmological parameters will be fixed to the 
best fit values of the WMAP data combined 
with the other sets of cosmological observables. This choice is not crucial since the values of the cosmological parameters are purely illustrative. 
The essential features of the present analysis are unchanged if the best fit values of the WMAP data alone (or partially combined 
with the other data sets) are consistently assumed.} 
 $ n_{\zeta} \simeq 0.947$ and $A(k_{\mathrm{p}}) \simeq 0.815$ when combined with the remaining cosmological data sets, i.e. supernovae and  large scale structure\footnote{The data 
\cite{HSTKP} and \cite{WL1,WL2} are not included in the joined analysis.}. 
In Eq. (\ref{b5}) the magnetic part of the correlation function 
is expressed as \cite{scalar2,scalar3}
\begin{equation}
{\mathcal P}_{\Omega}(k) = {\mathcal F}(\epsilon) \overline{\Omega}^2_{\mathrm{B\, L}}\biggl(\frac{k}{k_{\mathrm{L}}}\biggr)^{2 \epsilon},
\qquad {\mathcal F}(\epsilon)= \frac{4 ( 6 - \epsilon) ( 2 \pi)^{2\epsilon}}{\epsilon ( 3 - 2 \epsilon) \Gamma^{2}(\epsilon/2)},
\label{b6}
\end{equation}
where $2\epsilon$ the spectral index of the magnetic energy density 
fluctuations and $k_{\mathrm{L}}$ is the magnetic pivot scale 
that will be defined in a moment. In Eq. (\ref{b6}) we have that 
\begin{equation}
\overline{\Omega}_{\mathrm{BL}}= \frac{B_{\mathrm{L}}^2}{8\pi \rho_{\gamma}} = 7.56 \times 10^{-9} \biggl(\frac{B_{\mathrm{L}}}{\mathrm{nG}}\biggr)^2.
\label{b7}
\end{equation}
In Eq. (\ref{b7}) $B_{\mathrm{L}}$ is the value of the magnetic field 
smoothed, through a Gaussian window function,  over a typical comoving 
length $L$ which is related 
to the magnetic pivot scale as $L= 2\pi/k_{\mathrm{L}}$. In what 
follows the fiducial value $k_{\mathrm{L}} = \mathrm{Mpc}^{-1}$ 
will be consistently adopted. It is relevant to appreciate that $B_{\mathrm{L}}$ 
represents the smoothed magnetic field red-shifted to the present epoch.
This convention is a bit confusing but we will follow it since 
it is a common practice in the field. The source of the possible confusion 
is, in short, the following. The field $B_{\mathrm{L}}$ is the value of 
the magnetic field at recombination redshifted to the present epoch 
and assuming magnetic flux conservation which is well justified 
since the value of the conductivity is always rather large. However, 
${\mathrm B}_{\mathrm{L}}$ {\em is not}  the present value of the magnetic field intensity observed in galaxies and galaxy clusters. The rationale 
for this statement is that during the gravitational collapse of protogalaxies 
the magnetic field intensity $B_{\mathrm{L}}$ will be amplified by compressional amplification and, probably, also by some dynamo action.
Assuming just compressional amplification (which is the most certain 
aspect of the dynamics) the amplification factor may be of the order 
$10^{4}$ or even $10^{5}$. 

Notice that, from Eq. (\ref{b6}) the nearly scale-invariant limit is achieved 
for $\epsilon < 1$. Furthermore, as discussed elsewhere, the power spectrum 
of the magnetic energy density, being quartic in the field intensities, is computed in terms of the appropriate convolutions that have 
been evaluated, to get to Eq. (\ref{b6}), for $\epsilon< 1$. It should 
be borne in mind that, in the opposite case (i.e. $\epsilon > 1$) 
 the magnetic energy density has a violet spectrum. This implies that, at 
 very large length-scales the gravitational effect will be negligible and that, furthermore, the most significant bounds on the intensity of the magnetic field 
 will come from comoving momenta close to the diffusion scale. 
 Notice that the nearly scale-invariant spectrum is rather suggestive also 
 because it would imply a nearly scale-invariant spectrum of large-scale 
 magnetic fields from galaxies, to clusters, to super-clusters. This observation 
 is, today, beyond our observational capabilities. However, by looking at recent 
 determinations of magnetic fields in clusters and in some typical supercluster, it is indeed tempting to speculate that the resulting power spectrum 
 of the magnetic energy density is nearly scale-invariant. 
 
\begin{figure}
\begin{center}
\begin{tabular}{|c|c|}
      \hline
      \hbox{\epsfxsize = 7.6 cm  \epsffile{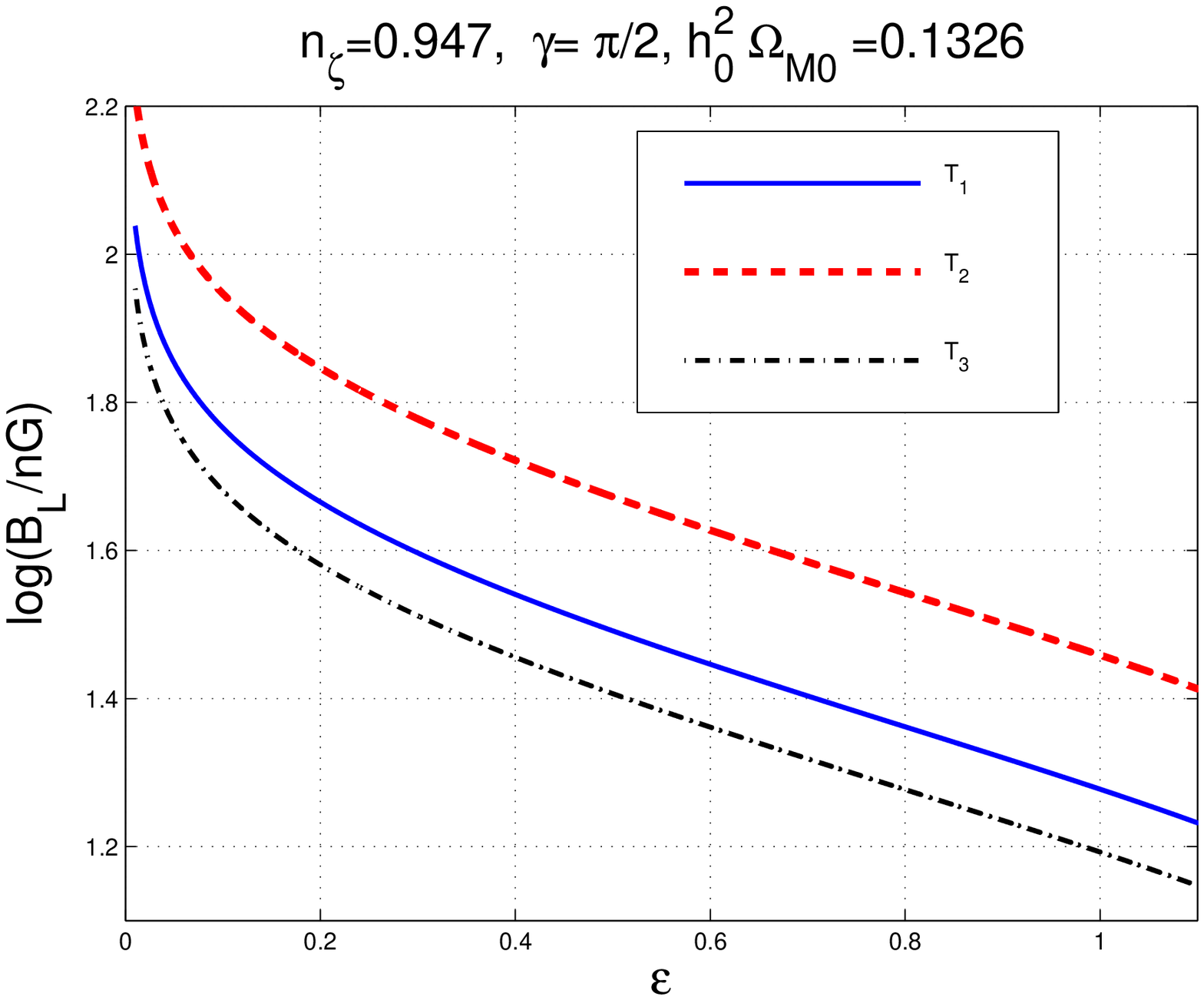}} &
      \hbox{\epsfxsize = 7.8 cm  \epsffile{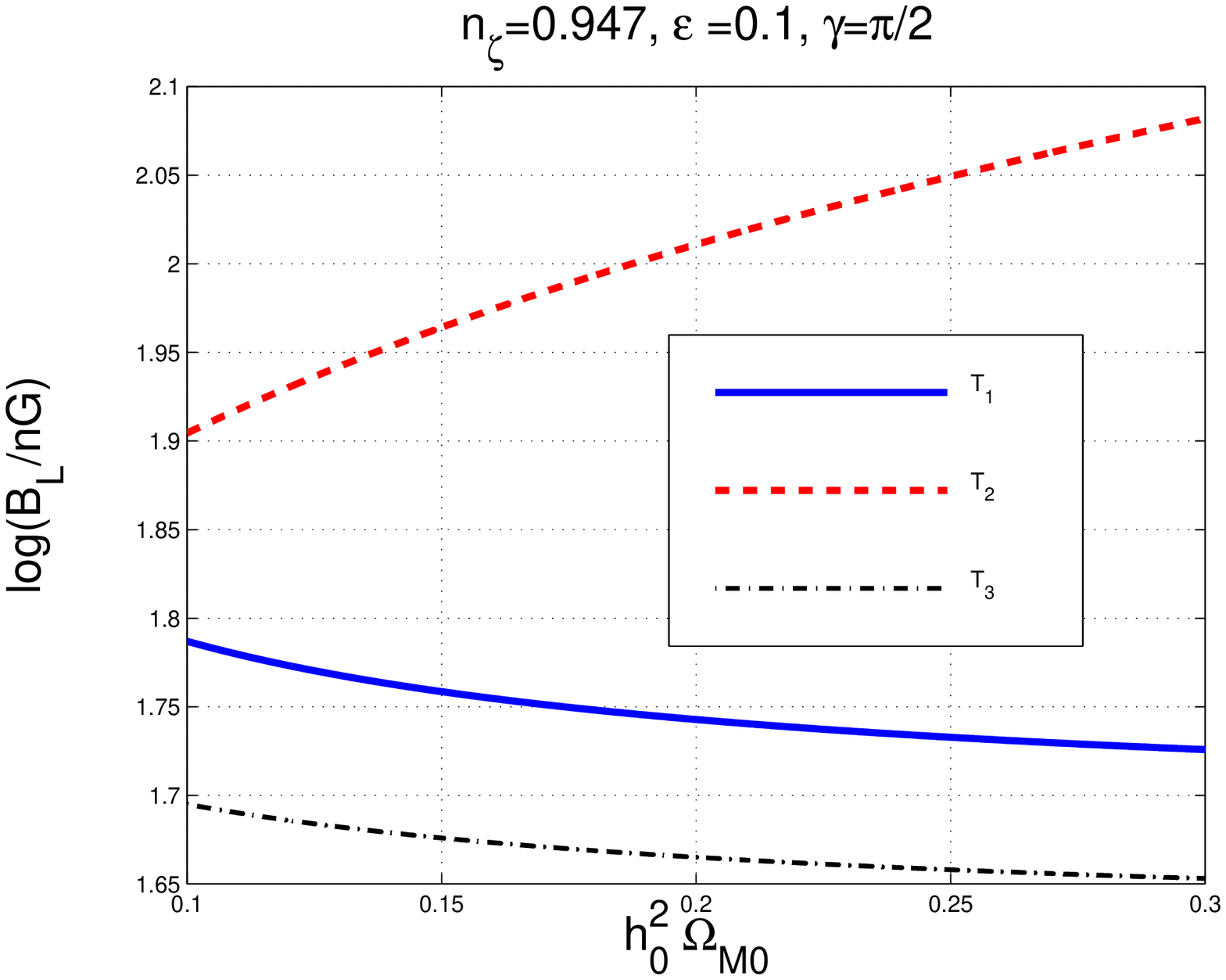}}\\
      \hline
\end{tabular}
\end{center}
\caption[a]{The bounds on the smoothed magnetic field intensity 
are illustrated as a function of the magnetic spectral index (plot at the left) 
and as a function of the critical fraction in dusty matter (plot at the right). The 
pivot scales $k_{\mathrm{p}}$ and $k_{\mathrm{L}}$ are fixed,
respectively, to $0.002\,\,{\mathrm Mpc}^{-1}$ and $\mathrm{Mpc}^{-1}$.}
\label{B1}
\end{figure}

The correlation function of curvature perturbations enters directly the 
determination of the ordinary Sachs-Wolfe contribution which is the dominant 
source of temperature anisotropies at large angular scales. 
Let us require that the magnetized contribution is always much smaller 
than the adiabatic contribution. So, we will assume, in a rather 
conservative approach, that the magnetized contribution 
is always smaller (by a factor $\eta$) than the adiabatic contribution. The parameter $\eta$ can be ${\mathcal O}(10^{-3})$, in a conservative 
approach.  Following this strategy the magnetic field 
intensity can be bounded in each thermal history of the Universe.
Consider, for instance, the plots appearing in Fig. (\ref{B1}) where, 
on the vertical axis, 
the base 10 logarithm of the smoothed magnetic field is reported 
in units of nG. In the plot at the left the three different lines 
correspond to the cases of the three different histories 
discussed in the previous sections. These three different 
histories will imply three different functions $T_{h}$ with $h = 1, 2,3$ and with, in general $T_1\neq T_{2}\neq T_{3}$. At late times the functions $T_{h}$ 
can be approximated as 
\begin{eqnarray}
&&T_{1}(\alpha_{\mathrm{rec}}, h_{0}^2 \Omega_{\mathrm{M}0}) \simeq 
\frac{3 R_{\gamma}}{4}\frac{\alpha_{\mathrm{rec}}}{3 \alpha_{\mathrm{rec}} + 4},
\label{b8}\\
&& T_{2}(\alpha_{\mathrm{rec}}, h_{0}^2 \Omega_{\mathrm{M}0}) 
\simeq \frac{R_{\gamma}}{3 \alpha_{\mathrm{rec}} + 4}, 
\label{b9}\\ 
&&T_{3}(\alpha_{\mathrm{rec}}, h_{0}^2 \Omega_{\mathrm{M}0})  \simeq \frac{R_{\gamma}}{12} \biggl( \frac{12 \alpha_{\mathrm{rec}} + 4}{3 \alpha_{\mathrm{rec}} + 4}\biggr).
\label{b10}
\end{eqnarray}
The function $T_{1}$ arises in the simplest case, i.e. 
when the Universe, after a phase of conventional inflation, passes from a 
radiation-dominated epoch to the matter stage. In $T_{1}$ the reheating 
is assumed to be a sudden process. The function $T_{2}$ describes 
the situation when there is an intermediate stiff phase expanding 
at a rate slower than radiation. Finally the function $T_{3}$ arises 
when there is a prolonged reheating and then the Universe 
is, subsequently, dominated by radiation and matter.
In Fig. \ref{B1} (plot at the left) we illustrate the bounds as a function of 
the magnetic spectral index. The region below each of the curves 
implies that the contribution of the magnetic field to the 
two-point function is smaller than $10^{-3}$. 

The less restrictive 
case is, surprisingly enough, $T_2$. The rationale for this occurrence 
can be simply understood in terms of the results reported in the previous 
section. In the transition from the stiff phase to the radiation phase 
the magnetized component gets reduced while in the subsequent 
transition from radiation to matter the magnetized contribution gets enhanced.
Since the two effects tend to cancel, the net result will be an overall 
suppression of the magnetized contribution. This dynamical 
suppression allows the magnetic field amplitude to be larger 
than in the other two cases (i.e. $T_1$ and $T_3$) where 
there is no "destructive" interference between the two subsequent 
effects.

In the plot at the right of Fig. \ref{B1} this pattern is confirmed with an extra
piece of information: if the matter fraction increases the bound 
become looser in the case of $T_{3}$ and a bit tighter in the case of $T_{1}$ 
and $T_{2}$. In fact, in Fig. \ref{B1} (plot at the right) the 
magnetic spectral index has been kept fixed while the matter fraction 
is allowed to move from the fiducial value $h_{0}^2 \Omega_{\mathrm{M}0}=0.1326$ which is the one assumed in the left plot of Fig. \ref{B1}.
The increase (or decrease) of the critical fraction of dusty matter simply means, physically, that the recombination epoch is either delayed or anticipated. 

It should also be remarked that the case of prolonged reheating is the 
most constraining. This occurrence can be also simply understood from the considerations 
of the previous section. In the case of prolonged reheating $c_{\mathrm{st}}^2 \neq 1/3$. So during this phase the curvature perturbations decrease and this 
decrease will interfere constructively with the usual radiation matter 
transition since both contribution have the same sign.
\begin{figure}
\begin{center}
\begin{tabular}{|c|c|}
      \hline
      \hbox{\epsfxsize = 7.6 cm  \epsffile{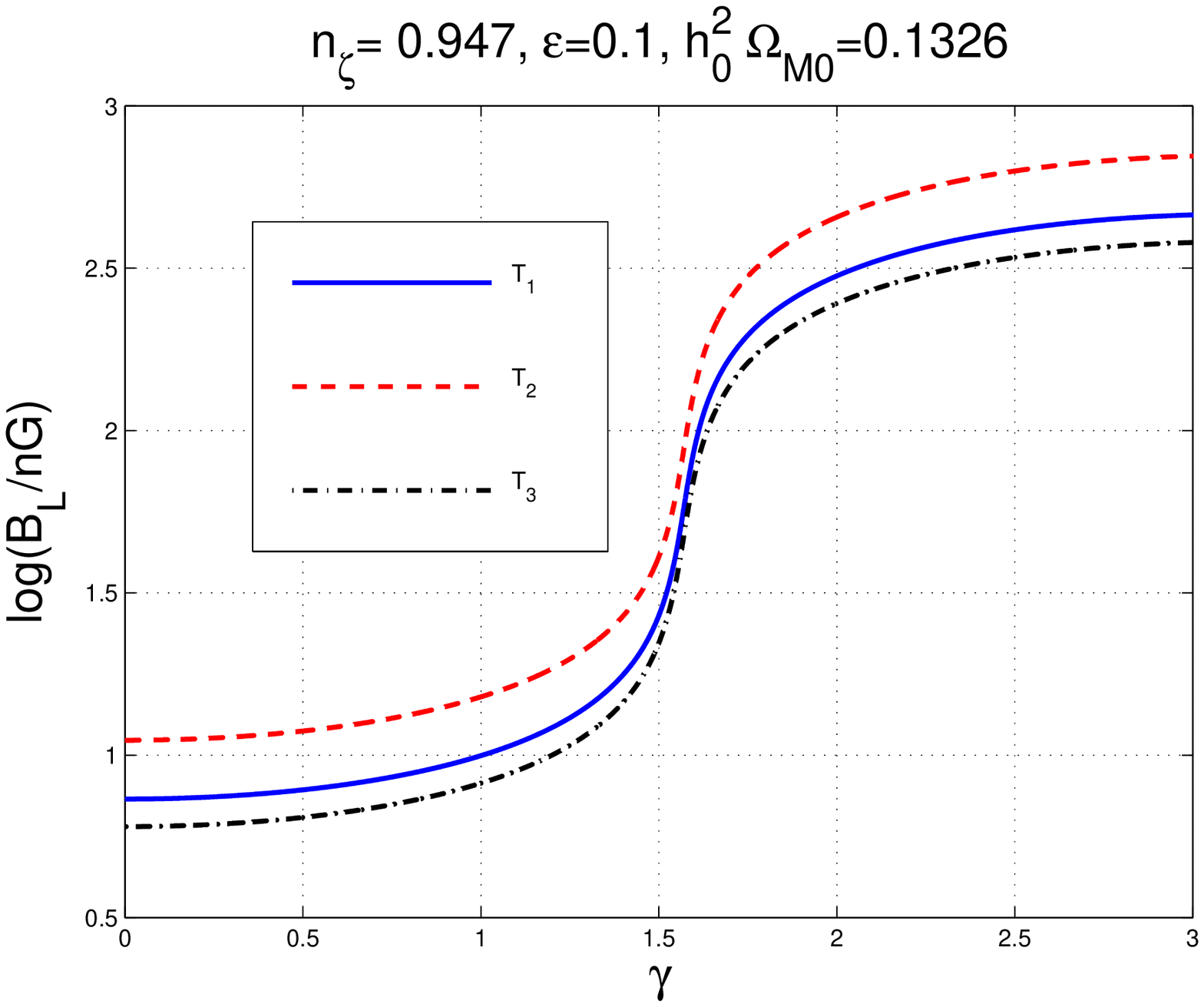}} &
      \hbox{\epsfxsize = 7.6 cm  \epsffile{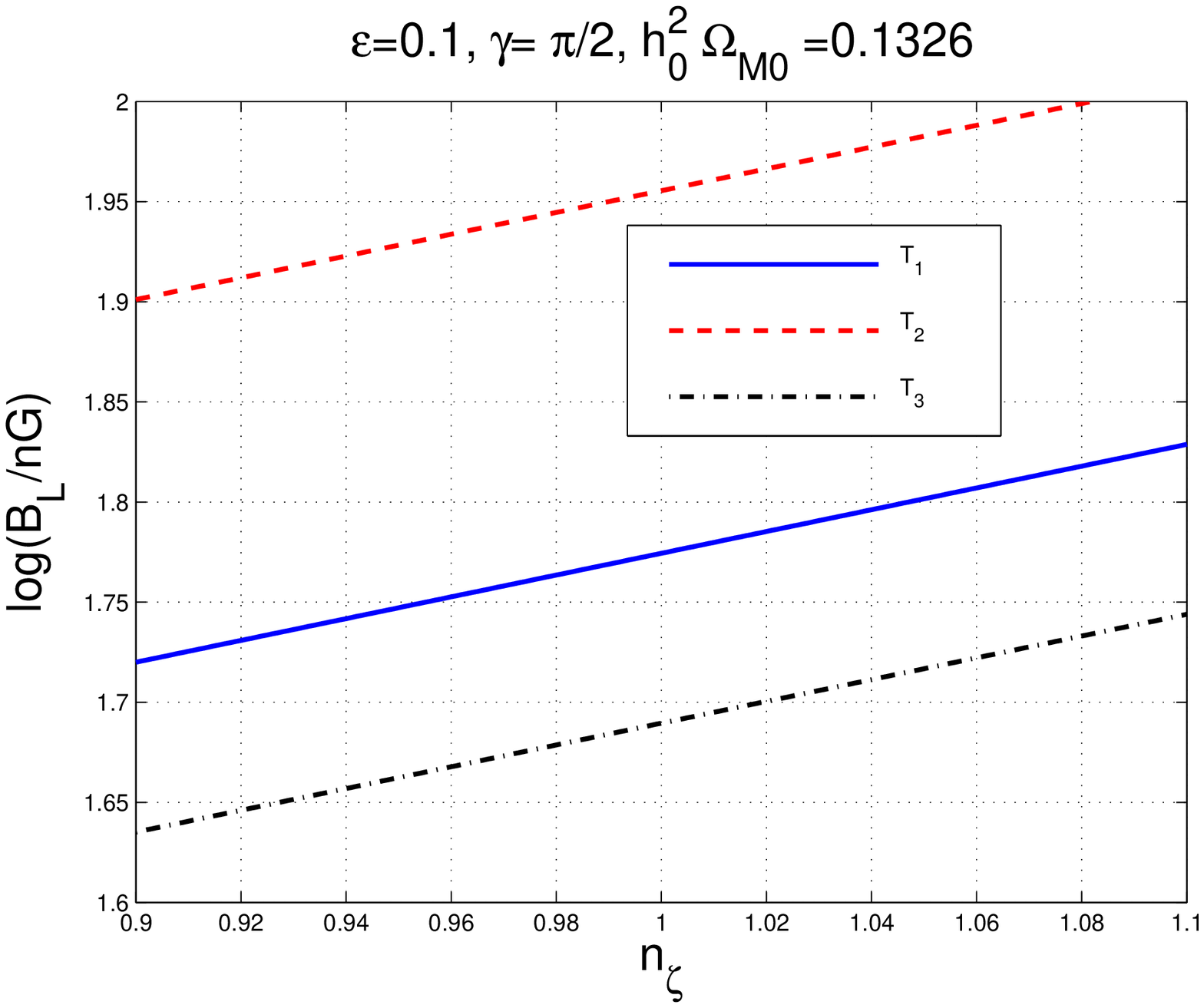}}\\
      \hline
\end{tabular}
\end{center}
\caption[a]{The bounds on the smoothed magnetic field intensity are illustrated as a function of the correlation angle $\gamma$ (plot at the left) and as a function of the adiabatic spectral index $n_{\zeta}$ (plot at the right). All the other parameters are kept fixed to their fiducial values.}
\label{B2}
\end{figure}
In Fig. \ref{B1} the value of the correlation angle has been fixed to $\pi/2$
implying that the adiabatic and the magnetized mode are uncorrelated.
Again we may relax this assumption and allow $\gamma$ to vary 
while all the other parameters are fixed to their fiducial values. This 
has been done in Fig. \ref{B2}. As expected the bounds is tighter in the 
case when the two components are correlated and looser when the two 
components are anti-correlated.  In Fig. \ref{B2} (plot at the right) 
the variation of the adiabatic spectral index is illustrated always in the 
case where the two components are uncorrelated.
\begin{figure}
\begin{center}
\begin{tabular}{|c|c|}
      \hline
      \hbox{\epsfxsize = 7.6 cm  \epsffile{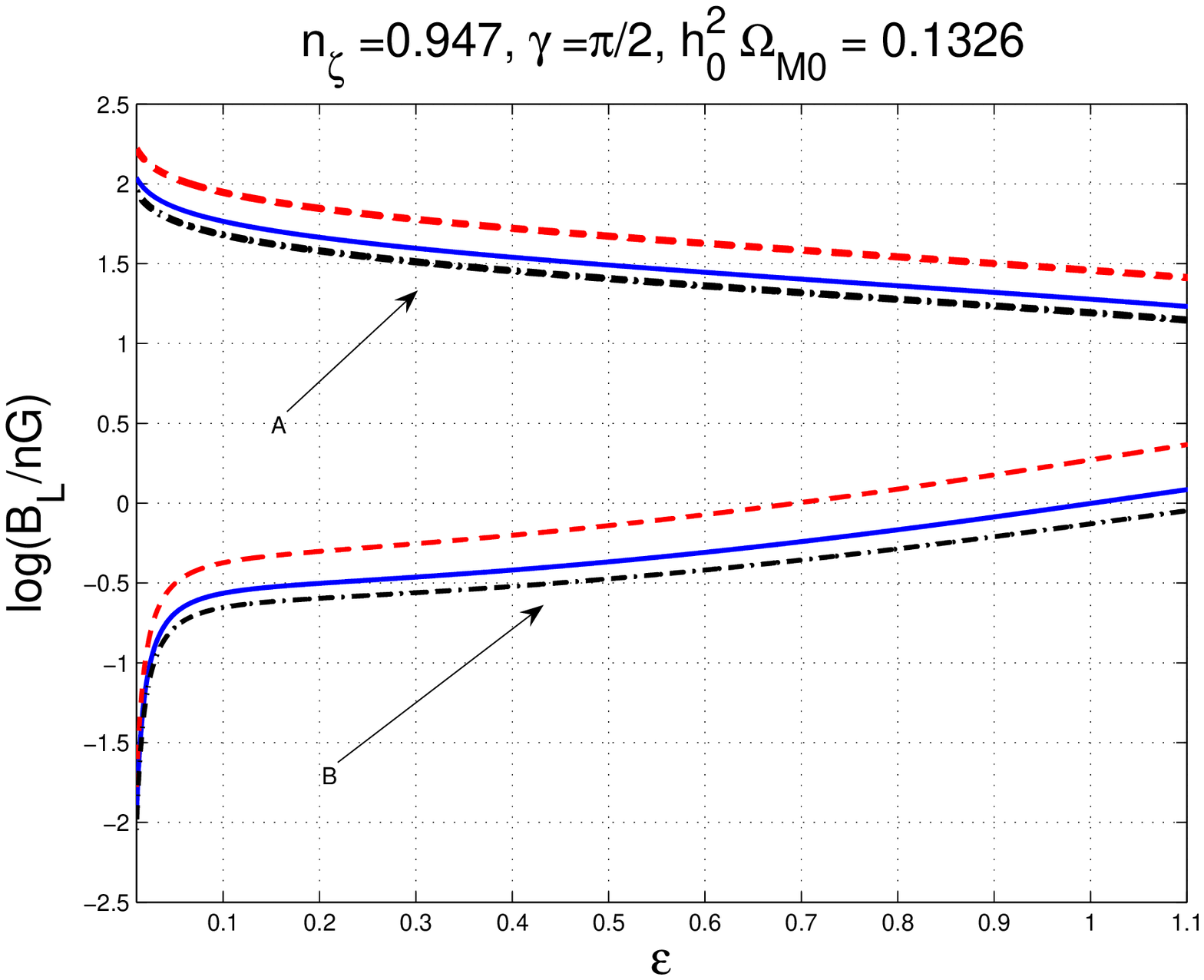}} &
      \hbox{\epsfxsize = 7.6 cm  \epsffile{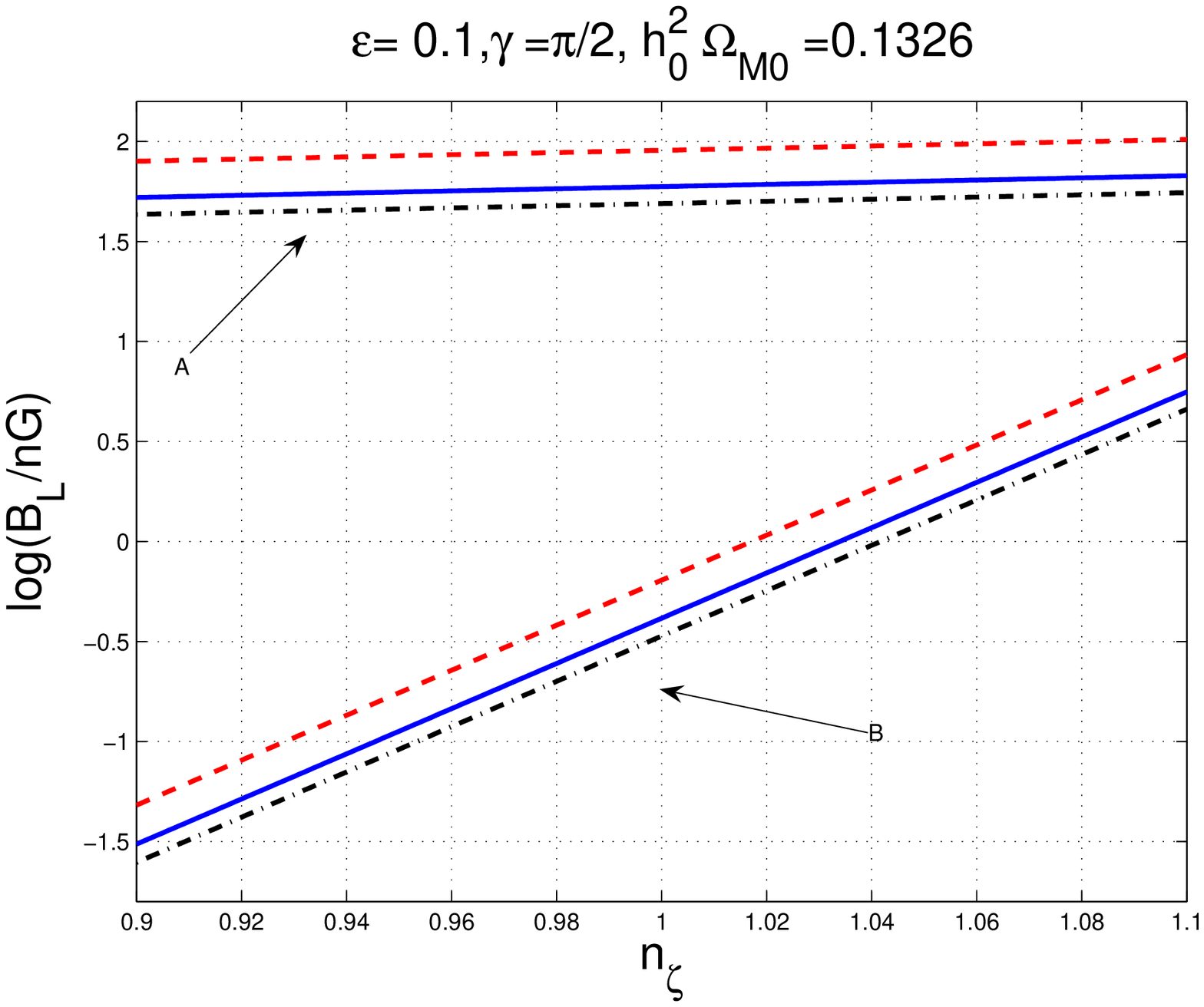}}\\
      \hline
\end{tabular}
\end{center}
\caption[a]{The potential differences in the estimate of the diffusion scale 
are illustrated. In the plot at the left, the bunch of curves labeled by B 
corresponds to the estimate of the diffusion scale given in Eq. (\ref{second}) 
while the curves labeled by A are the same illustrated in Fig. \ref{B1} (plot at the left). In the plot at the right the two classes of curves have illustrate
the same phenomenon but in terms of the dependence on $n_{\zeta}$ 
and should be compared with the right plot appearing in Fig. \ref{B2}.}
\label{B3}
\end{figure}
It is now relevant to remark that while $P_{\zeta}(k)$ slightly decreases 
as $k$ increases, the opposite is true for $P_{\Omega}(k)$. It is therefore 
clear that, given this situation, the most constraining wavenumbers are the 
largest, i.e. the small-scale behaviour of the spectrum is the one that 
leads to the most stringent bounds. This consideration has been 
taken into account in deriving the bounds expressed by Figs. \ref{B1} and 
\ref{B2}. In particular, it has been assumed that the diffusion scale is 
the one roughly associated with the Silk wavenumber, i.e. 
\begin{equation}
\frac{1}{ k_{\mathrm{D}} \tau_{\mathrm{rec}}} = 9.63\times 10^{-3} \biggl(\frac{h_{0}^2\Omega_{\mathrm{b}}}{0.023}\biggr)^{-1/2}
\biggl(\frac{h_{0}^2\Omega_{\mathrm{M}}}{0.134}\biggr)^{1/4} \biggl(\frac{1050}{z_{\mathrm{rec}}}\biggr)^{3/4}.
\label{first}
\end{equation}
There might be slightly different choices for the diffusion scale. For instance 
it has been noticed in the past that the diffusion scale of the magnetic fields 
should be related with the induced velocity of Alfv\'en waves. In this case 
the diffusion length-scale will be slightly smaller than the Silk scale by a 
factor which is essentially proportional, in our notations, to $(B_{\mathrm{L}}/{\mathrm{nG}})$. In this case the diffusion wavenumber will be given by
\cite{tensor1}: 
\begin{equation}
k_{\mathrm{D}} \simeq (1.7 \times 10^{2})^{2/\epsilon} \biggl(\frac{B_{\mathrm{L}}}{\mathrm{nG}}\biggr)^{- 2/(\epsilon +2)} h_{0}^{1/(\epsilon+2)} \,\, \mathrm{Mpc}^{-1}
\label{second}
\end{equation}
Figure \ref{B3} illustrates the situation described by Eq. (\ref{second}). It is 
clear that the patterns of the different thermal histories remain the same.
However, the bound is improved by roughly one order of magnitude.
\renewcommand{\theequation}{5.\arabic{equation}}
\setcounter{equation}{0}
\section{Concluding remarks}
\label{sec5}
There are various lessons that can be drawn from the exercise 
reported in this paper. The main question we addressed has been 
the possible influence of slight variations in the thermal history of the Universe 
on the curvature perturbations induced by a magnetic field present 
for typical scales that are larger than the Hubble radius at recombination.
Various inflationary mechanisms for the generation of cosmic magnetic field 
fall in this situation. It has been found that, indeed, the thermal history 
of the Universe may either suppress or increase the relative weight 
of magnetized curvature perturbations. The suppression occurs when, 
during the dynamical evolution, there is a sort of destructive interference.
This phenomenon takes place, amusingly enough, when there was a 
phase, in the life of the Universe, when the rate of expansion 
was slower than the one of radiation. This kind of evolution 
must occur prior to the onset of big-bang nucleosynthesis and explicit examples have been provided. Also the opposite 
phenomenon can be realized, i.e. a sort of constructive 
interference when the overall contribution of the magnetized curvature 
perturbations gets enhanced in comparison with the adiabatic contribution 
present at the end of inflation.  This situation takes place if the 
reheating phase is sufficiently prolonged in comparison 
with the sudden reheating approximation. 

These considerations can certainly be developed along different directions.
It is interesting to remark that the nature of the bounds on the 
magnetic field intensity may change quantitatively and qualitatively.
Different thermal histories imply that the constraints on the magnetic field 
intensities may vary, generally speaking, of one order of magnitude.
In the case when the Universe contains after inflation (but before radiation)
a stiff phase the present value of the smoothed magnetic field 
must be qualitatively\footnote{It should be stressed, as discussed in section 4, that $B_{\mathrm{L}}$ is the intensity of the magnetic field at recombination redshifted to the present epoch. This field is by no means equal to the present magnetic field. So the bounds on $B_{\mathrm{L}}$ are, really and truly, 
bounds on a primordial magnetized background. The present magnetic field of galaxies and clusters is certainly related to $B_{\mathrm{L}}$. In the simplest 
case, compressional amplification may turn a 0.1 nG field (coherent 
over a comoving scale of the order of the Mpc) into the $\mu$G 
field observed in galaxies.} $B_{\mathrm{L}} < 1.5\times 10^{-7}$G
while in the context of prolonged reheating it must be as small as
$1.6 \times 10^{-8}$ G. 

The presented considerations are not totally insensitive 
on the estimate of the diffusive wavenumber. In particular, it occurs that 
while experimental data favour a slightly red spectrum of curvature 
perturbations (i.e. an adiabatic spectral index slightly smaller than $1$), the 
magnetized contribution is typically nearly scale-invariant but slightly blue.
In this situation the most constraining wavenumbers are the ones 
close to diffusion scale. If the diffusivity length-scale is slightly smaller than the Silk scale, then the overall bounds on the smoothed magnetic field 
intensity range, depending on the details of the specific thermal history, 
from $B_{\mathrm{L}}< \mathrm{nG}$ to $B_{\mathrm{L}}< 0.1 $ nG. 
It can be discussed if the Alfv\'en velocity should enter 
the thermal diffusivity scale when the magnetic field is fully inhomogeneous.
Indeed Alfv\'en waves are only excited when there is a background magnetic field in the game. This is not the case when the magnetic field 
does not break spatial isotropy which the case investigated in the present paper and which is also, in our opinion, the most realistic one (see the introduction). With these caveats, however, it is important to notice 
that different thermal histories may either relax or improve the bounds on the magnetic field intensity by one order of magnitude.

\end{document}